\documentstyle[12pt]{article}

\begin{document}
\begin{center}
\large {\bf COSMOLOGICAL CONSTANT, CONICAL DEFECT AND CLASSICAL TESTS
OF GENERAL RELATIVITY}
\end{center}
\centerline{Wilson H. C. Freire $^{1,2}$, V. B. Bezerra $^{2}$,
 J. A. S. Lima $^{3}$}

\begin{center}
$^1$ Universidade Regional do Cariri, \\Departamento de Matem\'atica\\
63100-000 Crato, Ce, Brazil.
\end{center}

\begin{center}
$^2$Universidade Federal da Para\'{\i}ba, \\Departamento de
F\'{\i}sica, Caixa Postal 5008, \\58059-970 J. Pessoa, Pb, Brazil.
\end{center}

\begin{center}
$^3$Universidade Federal do Rio Grande do Norte,
\\Departamento de F\'{\i}sica, Caixa Postal 1641, \\59072-970 Natal, RN,
Brazil.
\end{center}

\vskip 0.6cm

\centerline{Abstract}
\noindent

We investigate the perihelion shift of the planetary motion
and the bending of starlight in the Schwarzschild field modified by the
presence of a $\Lambda$-term plus a conical defect. This analysis generalizes an earlier result obtained by Islam (Phys. Lett. A {\bf 97},
239, 1983) to the case of a pure cosmological constant. By using the
experimental data we obtain that the parameter $\epsilon$ characterizing the
conical defect is less than $10^{-9}$ and $10^{-7}$, respectively,
on the length scales associated with such phenomena. In particular, if the
defect is generated by a cosmic string, these values correspond to limits on
the linear mass densities of $10^{19}g/cm$ and $10^{21}g/cm$, respectively.
\newpage
\noindent

The best theoretical framework for describing the gravitational interaction
is provided by the general relativity theory (GRT). The major achievements
of the theory, namely, the deflection of starlight and the perihelium shift
of the Mercury planet in the Sun field, agree with the Einstein values
with an accuracy of one percent (an overview is given by Will\cite{CW 85}).
As widely known, these successful analyses have been carried out in the
context of the original spherically symmetric Schwarzschild line element.

Possible modifications of the Schwarzschild spacetime may either preserve
spherical symmetry or depart slightly from it. In the former case one may
include several contributions, among them: a cosmological $\Lambda$-term
(Kottler metric\cite{KO 18}), a net electric charge (Reissner-Nordstr\"om
solution\cite{R 16,N 18}), and the presence of a magnetic
monopole\cite{RA 76}. At present, the last two possibilities do not appear
to be interesting in physical grounds, however, the case for a cosmological
constant is still important, because of its connection with possible
gravitational effects of the vacuum energy density in  black hole
physics\cite{BH}, as well as in the cosmological context\cite{KT 95}.

The influence of such effects in the so-called classical tests are usually
analysed assuming  negligible contributions from solar oblatness or whatever
effect departing the metric from exact spherical symmetry, e.g., due to the
presence of a conical defect, which is commonly exemplified by a cosmic
string. Naturally, if the Sun deviates slightly from exact spherical
symmetry, either due to an appreciable solar quadrupole moment\cite{DG 67}
or even some unexpected topological property of the gravitational field
(matter distribution), a more complete treatment of these effects require
a generalization of the spherically symmetric form of the line element. In
the case of a nonzero small quadrupole moment, the main physical consequences
have been discussed with some detail in the literature either in the
Newtonian approximation or in the relativistic framework(see\cite{IG 66}
and refs. therein). However, as far as we know, an exact treatment including
a conical defect has not been considered in the literature, and its influence
is not only matter of academic interest, because such a possibility
basically remains as an open question.

On the other hand, some authors have suggested that the most simple
exact solutions of Einstein's equations can easily be generalized to
include a conical defect\cite{VS 94}. Such spacetimes are
geometrically constructed by removing a wedge, that is, by requiring
that the azimuthal angle around the axis runs over the range $0 < \phi <
2\pi b$. For very small effects the parameter $b$ itself may be written as
 $b = 1 - \epsilon$, where $\epsilon$ is a small dimensionless parameter
quantifying the conical defect. In particular, for $\epsilon = 0$
the spherically symmetric line element is recovered whereas for a conical
defect generated by a cosmic string one has $\epsilon = {8G\mu/c^2}$,
where $\mu$ is the mass per unit length of the string\cite{VI 81,WH 85}.

In this paper, we study the orbits of massive and massless particles in
the gravitational field of the sun modified by a conical defect. As we shall
see, our simplified analysis of the perihelion shift and the bending of
light, provide two upper limits on the value of the conical defect
parameter $\epsilon$. For completeness, in our calculations we have also
included the presence of a cosmological constant, thereby generalizing the results previously derived by Islam\cite{JN 83}.

The Kottler spacetime endowed  with a conical defect takes the
following form

\begin{eqnarray*}
\label{eq1}
ds^2=\left(1-\frac{2M}{r} + \frac{\Lambda r^2 }{3} \right)c^2dt^2 -
\left(1 - \frac{2M}{r} +\frac{\Lambda r^2 }{3} \right)^{-1}dr^2
 \end{eqnarray*}
\begin{equation}
 - r^2 d\theta^2 - b^2r^2{\sin^{2}}\theta d\phi^{2} ,
\end{equation}
where $M= Gm/c^2$ is the geometric mass of the central body, $\Lambda$ is
the cosmological constant, and b is the conical defect parameter.

In that spacetime, a test particle follows
geodesic equations which can be obtained from the Lagrangian

\begin{eqnarray*}
\label{eq2}
L = \frac{1}{2}\left[\left(1 -\frac{2M}{r}+\frac{\Lambda r^2}{3} \right)
( \frac{cdt}{dp})^2 - \left(1 -\frac{2M}{r}+
    \frac{\Lambda r^2}{3} \right)^{-1}
     ({\frac{dr}{dp}}) ^2\right.
\end{eqnarray*}
\begin{equation}
\left.- r^2 (\frac{d\theta}{dp})^2 - b^2 r^2\sin^2 \theta
 (\frac{d\phi}{dp})^2 \right] ,
\end{equation}
where $p$ is an affine parameter.

Using the Euler-Lagrange approach, the equations of motion read

\begin{equation}
\label{eq3}
\frac{d}{dp}\left[(1 -\frac{2M}{r} + \frac{\Lambda r^2}{3})\frac{dt}
{dp}\right]=0,
\end{equation}

\begin{equation}
\label{eq4}
\frac{d}{dp}(r^2 \frac{d\theta}{dp}) - b^2 r^2 \sin\theta \cos\theta
 (\frac{d\phi}{dp})^2 =0,
\end{equation}

\begin{equation}
\label{eq5}
\frac{d}{dp}(b^2 r^2 \sin^2 \theta \frac{d\phi}{dp})=0.
\end{equation}

The first geodesic equation give us the following result
\begin{equation}
\label{eq6}
\frac{dt}{dp} = \frac{E}{c}\left(1 -\frac{2M}{r} +
\frac{\Lambda r^2}{3} \right)^{-1},
\end{equation}
where $E$ is a constant. Note that in the Newtonian limit $E$ corresponds
to the relativistic energy of the particle.

For simplicity, let us choose the initial conditions $\theta =\pi /2$ and
$\frac{d\theta}{dp} = 0$. Then, from Eq.(\ref{eq4}) we find that
$\frac{d^2{\theta}}{dp^2} = 0$. This means that the motion is confined to
the plane $\theta = \pi/2$, and this fact allow us to simplify all the
remaining equations inserting everywhere $\theta = \pi/2$. Therefore,
Eq.(\ref{eq5}) implies that

\begin{equation}
\label{eq7}
r^2 \dot{\phi}=L,
\end{equation}
where $L$ is a  constant of motion, and a dot means derivative with respect to affine parameter.

In order to get the solution for the radial coordinate, let us now consider
the standard constraint equation

\begin{equation}
\label{eq8}
g_{\mu\nu}{\frac{d x^\mu}{dp}}{\frac{d x^\nu}{dp}} = \kappa,
\end{equation}
where $\kappa$ is a constant, and we can fix the parameter $p$ by taking
$\kappa = - c^2$, \\ $c^2$  or $0$, for space-like, time-like or
light-like curves. Therefore, for the radial equation we obtain the following
result

\begin{equation}
\label{eq9}
\left(1 -\frac{2M}{r} + \frac{\Lambda r^2}{3} \right)c^2 (\frac{dt}{dp})^2
 -\left(1 -\frac{2M}{r} + \frac{\Lambda r^2}{3}
   \right)^{-1}(\frac{dr}{dp})^2 - b^{2}r^2 (\frac{d\phi}{dp})^2
       = \kappa.
\end{equation}

Now, replacing (\ref{eq6}) and (\ref{eq7}) into (\ref{eq8}), we get

\begin{equation}
\label{eq10}
\frac{1}{2}(\frac{dr}{dp})^2 + V(r) = \frac{1}{2}E^2,
\end{equation}
where the effective potential is $V(r) =  \frac{1}{2}\kappa +
\frac{\Lambda b^{2} L^{2}}{6} - \frac{\kappa M}{r}
+ \frac{b^2 L^2}{2r^2} - \frac{Mb^2 L^2}{r^3} +
\kappa \frac{\Lambda r^2}{6}$.

Let us now determine the orbits. First of all, we change the variable $r$
to $u=r^{-1}$ so that $\frac{dr}{dp} = -L \frac{du}
{d\phi}$. Then, for non-circular orbits and massive
particles $(\kappa = c^2)$ equation (\ref{eq9}) becomes

\begin{equation}
\label{eq11}
\frac{d^2 u}{d\phi ^2} + b^2 u =\frac{c^2 M}{L^2} + 3M b^2 u^2 +
\frac{c^{2} \Lambda}{3 L^{2}u^3} ,
\end{equation}
which reduces to equation (13) of Islam's paper\cite{JN 83} in the limiting case b=1 (note that in his notation $L^{2}/c^2 = b^2$).

The first term on the right hand side (RHS) of (\ref{eq11}) leads to quasi-
Newtonian  orbits. In fact, if we consider only this term into (\ref{eq11}),
the solution is given by the result

\begin{equation}
\label{eq12}
u_{0}=\frac{1}{r}=\frac{c^2 M}{b^2 L^2}\left[1 + e\cos(b(\phi -
   \phi _{0}))\right],
\end{equation}
where $\phi_0$ and $e$ are constants of integration (in this form $e$ is the eccentricity of the orbit). Notice that for $b=1$ we obtain the Newtonian
result.

In principle, in order to obtain the full corrections to the Newtonian orbits, equation (\ref{eq11}) should be exactly integrated. However, in order to compare the results with the astronomical observations, the simplest way is provided by the method of sucessive aproximation. The first order correction may easily be obtained by considering the
perturbative expansion, $u\cong u_{0}+u_{1}$ ($u_{1} << u_{0}$), where $u_{0}$ is given by (\ref{eq12}). The application of the standard perturbative procedure to this extended framework is justifiable because the last two terms in (\ref{eq11}) are small in comparison to the Newtonian contribution. This can be checked by comparing the relative magnitudes of each term using that (i) the conical defect $b$ is smaller than unity, and (ii) the extreme smallness of $\Lambda$ as provided by the present day cosmological limits\cite{P98}.

Now, considering orbits of small exccentricity we obtain the following equation for $u_1(\phi)$

\begin{equation}
\label{eq13}
\frac{d^2 u_{1}}{d\phi ^2} + b^2 u_{1}\cong \frac{6M^3 c^4}{b^2 L^4}
 \left(1 - \frac{b^{8} L^{8} \Lambda}{6c^{8} M^6} \right)
ecos[b(\phi -\phi _{0})],
\end{equation}
whose  solution is given by

\begin{equation}
\label{eq14}
u_{1}= \frac{3M^3 c^4}{b^4 L^4}\left(1 - \frac{b^8 L^8 \Lambda}{6c^8 M^6}
    \right)eb\phi sin[b(\phi -\phi _{0})].
\end{equation}

Including this correction, we have that

\begin{equation}
\label{eq15}
u\cong  \frac{c^2 M}{b^2 L^2}\left\{1 + e\left[\cos[b(\phi -\phi _{0})]
    + \frac{3M^2 c^2}{b^2 L^2}\left(1 - \frac{b^{8} L^{8} \Lambda}
{6c^{8}M^6} \right) b\phi sin[b(\phi -\phi _{0})]\right]\right\},
\end{equation}
and defining

\begin{equation}
\label{eq16}
\bigtriangleup \phi _{0}= 3\left(\frac{c M}{bL}\right)^2\left(1 -
 \frac{b^{8} L^{8} \Lambda}{6c^{8} M^6} \right) \phi,
\end{equation}
the solution $u(\phi)$ may be written as

\begin{equation}
\label{eq17}
u=\frac{1}{r}\cong \frac{c^2 M}{b^2 L^2}\left\{1 + e\cos[b(\phi -
   \phi_{0}- \bigtriangleup \phi_{0})]\right\}.
\end{equation}

It thus follows from (\ref{eq17}) that the required shift per revolution is

\begin{equation}
\label{eq18}
\bigtriangleup \phi = {\frac{6\pi}{b^3}}\left(\frac{cM}{L}\right)^2
\left(1 - \frac{b^{8} L^{8} \Lambda}{6c^{8}M^6} \right)
 + 2\pi(\frac{1}{b} - 1).
\end{equation}

By considering $b=1$ and using the fact that for the planet Mercury the
perihelion shift is determined with an accuracy better than
$5$x$10^{-3}$\cite{CW 85}, we find that

\begin{equation}
\label{eq18b}
|\Lambda| < 10^{-42} cm^{-2} ,
\end{equation}
a result previously obtained by Islam\cite{JN 83}

Let us now analyze the opposite limit ($\Lambda = 0$). As one may check,
expanding the resulting expression to first order in the conicity parameter
we obtain

\begin{equation}
\label{eq19}
\Delta\phi - \Delta\phi_S \approx 2\pi\epsilon,
\end{equation}
where $\Delta\phi_S$ is the standard deviation in the Schwarzschild
field. Recalling again that $\Delta\phi$ agree
with $\Delta\phi_S$ to better than $5$x$10^{-3}$, in order
to concilliate theory and observation, the parameter of conicity must be
bounded by

\begin{equation}
\label{eq20}
\epsilon < 10^{-9}
\end{equation}

In particular, this limit implies that if the defect is associated with a
cosmic string, its linear mass density $\mu$ is such that $\mu < 10^{19}g/cm$.
However, we stress that our results are completely general in the sense that
any deviation of the $b$ parameter from unity can be associated with a
nontrivial topology, not necessarily produced by a cosmic string.
In addition, if some portion of the perihelion shift
is due to other effects, like the quadrupole momentum of the sun, this
upper limit would be modified by the corresponding amount. In principle,
better limits should be available from Icarus and outer planets, however
the data for these cases are much less precise.

In order to determine the bending of a light ray  incoming from infinity
and passing near the Sun with impact parameter $D$, we need only to
consider $\kappa = 0$
in (\ref{eq9}). It thus follows that

\begin{equation}
\label{eq21}
\frac{d^{2}u}{d\phi ^{2}} + b^2 u = 3Mb^2 u^2.
\end{equation}

Therefore, different of what happens with the cosmological constant, we see
that the conical defect modify the light path (in this connection
see\cite{JN 83}). Proceeding in analogy with the perihelion shift we find
that the general solution of this equation is

\begin{equation}
\label{eq22}
u=\frac{1}{r} = \frac{1}{D}\sin(b\phi)+ \frac{1}{2}\frac{M}{D^2}
  [3 -4\cos(b\phi)+ \cos(2b\phi)].
\end{equation}

For $\phi = \pi + \delta \phi$, with $\delta\phi$ small, we have

\begin{equation}
\label{eq23}
u\cong \frac{4M}{D^2} - \frac{b\delta \phi}{D} + {\frac{\pi}{D}}(1 - b).
\end{equation}

Therefore the angular shift of the light ray is given by

\begin{equation}
\label{eq24}
 \delta\phi = \frac{4M}{bD} + \pi( \frac{1}{b} - 1).
\end{equation}

 Expanding this expression up to the first order in $\epsilon$, we get

\begin{equation}
\label{eq19b}
\delta\phi - \delta\phi_S \approx \pi\epsilon.
\end{equation}

 Measurements of deflection using long base line interferometric
 techniques for radio waves emmited by quasars are much less scattered,
 usually ranging from 1.57'' to 1.82'' with a precision of about 0.2''.
 As matter of fact, there are some expectations that such a error can
 even be reduced at least one order of magnitude, thereby providing an
 accurate test of general relativity\cite{RS 71,FS 76}. Assuming a rather
 conservative point of view, if $\delta \phi$ agree with  $\delta
\phi_S$ better than $0.31"$, that is, $\delta \phi = 1.75" \pm 0.31"$,
 we find from $(\ref{eq19b})$ that at the solar system length scale the
 conical defect parameter is bounded by

\begin{equation}
\label{eq19c}
\epsilon < 10^{-8}.
\end{equation}

Since the curvature of the Schwarzschild field with a conical defect
does not depend on $\epsilon$, such effects are uniquely
due to topological features or equivalently, due to the lack of
spherical symmetry produced by the conical defect. However, if we assume
that the conical defect is associated with a cosmic string,
then this limit give us a bound for $\mu$ which is two orders of magnitude
larger than the corresponding established by the perielion shift, that is,
$\mu$ is less than $10^{21} g/cm$ which is in agreement with the mass for GUT scale strings.

\centerline{\bf Acknowledgments}

 It is a pleasure to thank Robert Brandenberger and A. R. Plastino
 for a critical reading of the manuscript. This work is partially
 supported by the project Pronex/FINEP (No. 41.96.0908.00) and
 Conselho Nacional de Desenvolvimento Cient\'{i}fico e Tecnol\'{o}gico-
 CNPq (Brazilian Research Agency).


\begin{thebibliography}{99}
\bibitem{CW 85} C. M. Will {\it Theory and Experiment in Gravitational
Physics}, Cambridge University Press, Cambridge (1985).
\bibitem{KO 18} F. Kottler, {\it Ann. der Phys.} {\bf 56}, 410 (1918).
\bibitem{R 16} H. Reissner, {\it Ann. der Phys.} {\bf 50}, 106 (1916).
\bibitem{N 18} G. Nordstr\"om, {\it Proc. Kon. Ned. Akad. Wet.} {\bf 20},
1238 (1918).
\bibitem{RA 76} R. Adler, {\it Phys.Rev.} {\bf D 14}, 392 (1976).
\bibitem{BH} S. A. Hayward, T. Shiromizu and K. Nakao,  {\it Phys. Rev. D},
{\bf 49}, 5080 (1994).
\bibitem{KT 95} L.M. Krauss and M.S. Turner, {\it Gen. Rel. Grav.}, {\bf 27},
1137 (1995).
\bibitem{DG 67} R. H. Dicke and H. M. Goldberg, { \it Phys. Rev. Lett.},
{\bf 18}, 313 (1987).
\bibitem{IG 66} I. Goldberg, {\it Phys. Rev. Lett.}, {\bf 149}, 1010 (1966).
\bibitem{VS 94} A. Vilenkin and E. P. S. Shellard, {\it Cosmic Strings and
other Topological Defects}, Cambridge University Press (1994).
\bibitem{VI 81} A. Vilenkin, {\it Phys. Rev. D}, {\bf 23}, 852 (1981).
\bibitem{WH 85} W. A. Hiscock, {\it Phys. Rev. D}, {\bf 31}, 3288
(1985); J. R. Gott, {\it ApJ}, {\bf 288},422 (1985).
\bibitem{JN 83} J. N. Islam, {\it Phys. Lett. A}, {\bf 97}, 239(1983).
\bibitem{RS 71} R. A. Sramek, {\it Ap. J.}, {\bf 167}, L55 (1971).
\bibitem{FS 76} E. B. Fomalont and R. A. Sramek {\it Phys. Rev. Lett.},
{\bf 36}, 1475 (1976).
\bibitem{P98} Perlmutter, S. et al., Nature, 391, 51 (1998); Riess, A. G. et al. Astron. J., 116, 1009 (1998).

\end{thebibliography}
\end{document}